\documentclass[]{article}
\usepackage[utf8]{inputenc}
\usepackage{amsmath}
\usepackage[pdftex]{graphicx}
\usepackage{amscd}
\usepackage{calrsfs}
\usepackage{slashed}
\usepackage{amsfonts}
\usepackage{authblk}

\title{Complex ER bridges in EPRB decays}
\author{Oscar Brauer and Miguel Socolovsky}
\affil{Instituto de Ciencias Nucleares \\ Universidad Nacional Aut\'onoma de M\'exico \\e-mails: brauer@ciencias.unam.mx, socolovs@nucleares.unam.mx}
\providecommand{\keywords}[1]
{
	\small	
	\textbf{Keywords:} #1
}
\begin{document}
 \date{}
\maketitle

\begin{abstract}
	We argue that a great circle in the 7-sphere plays the role of an Einstein-Rosen bridge in Einstein-Podolsky-Rosen-Bohm decays.

\end{abstract}

\section{Introduction}

In recent years, it has been pointed out [1] that even in simple quantum mechanical entangled systems like Einstein-Podolsky-Rosen-Bohm (EPRB) [2,3] decaying processes producing pairs of spin ${{1}\over{2}}$ particles or photons, an Einstein-Rosen bridge or wormhole between the decay products exists. However, except for qualitative arguments in favour of this idea, no explicite example for this fact exists in the literature.

In this note, for the case of an entangled pair of i) two spin ${{1}\over{2}}$ massive particles flying (non relativistically) appart from each other, product of the decay of an initial particle in a singlet spin state, and ii) two photons e.g. in positronium decay $e^+e^-\to \gamma\gamma$ (odd parity) [4] or in radiative cascade of calcium $J=0\to J=1\to J=0$ (even parity) [5], we show that the geometrical object which plays the role of a \textit{bridge between the decaying particles} is a \textit{great circle of the unit 7-sphere} ($S^7$), geodesic [6] of the Fubini-Study metric [7] in the 3-dimensional complex projective space ($\mathbb{C}P^3$). This great circle, however, does not live in ordinary space or spacetime, but in the complex space $\mathbb{C}^4\cong\mathbb{C}^2\otimes\mathbb{C}^2$, which is the Hilbert space of the system.

\section{Non relativistic EPRB decay}
The initial normalized state of two massive spin ${{1}\over{2}}$ particles with magnetic moment $\mu$ flying apart from each other in the \textit{y}-direction, decay products of an initial particle with total zero spin, is the entangled pure state
\begin{equation}   
\begin{split}
|\psi(t=0)>&={{1}\over{\sqrt{2}}}(|\uparrow_1>\otimes|\downarrow_2>-|\downarrow_1>\otimes|\uparrow_2>)\\&\cong {{1}\over{\sqrt{2}}}\left( \left(^{1}_{0} \right)  _1\otimes\left(^{0}_{1}\right)_2-\left( ^{0}_{1} \right) _1\otimes \left( ^{1} _{0} \right) _2 \right) \\&\cong 
\begin{pmatrix}
0\\
\dfrac{1}{\sqrt{2}}\\
-\dfrac{1}{\sqrt{2}}\\
0
\end{pmatrix}
\end{split}
\end{equation}
where we have chosen to make the spin measurements in the \textit{z}-direction with two Stern-Gerlach (S.G.) apparata with exterior magnetic field $B_z=B_0+\alpha z$, $\alpha=const.>0$. (See Fig. 1, where $x,y,z$ are ordinary space coordinates, and $t$ is the time interval that each of the particles spends in the corresponding S.G.) 

\begin{figure}[h]
	\centering
	\includegraphics[width=1\linewidth]{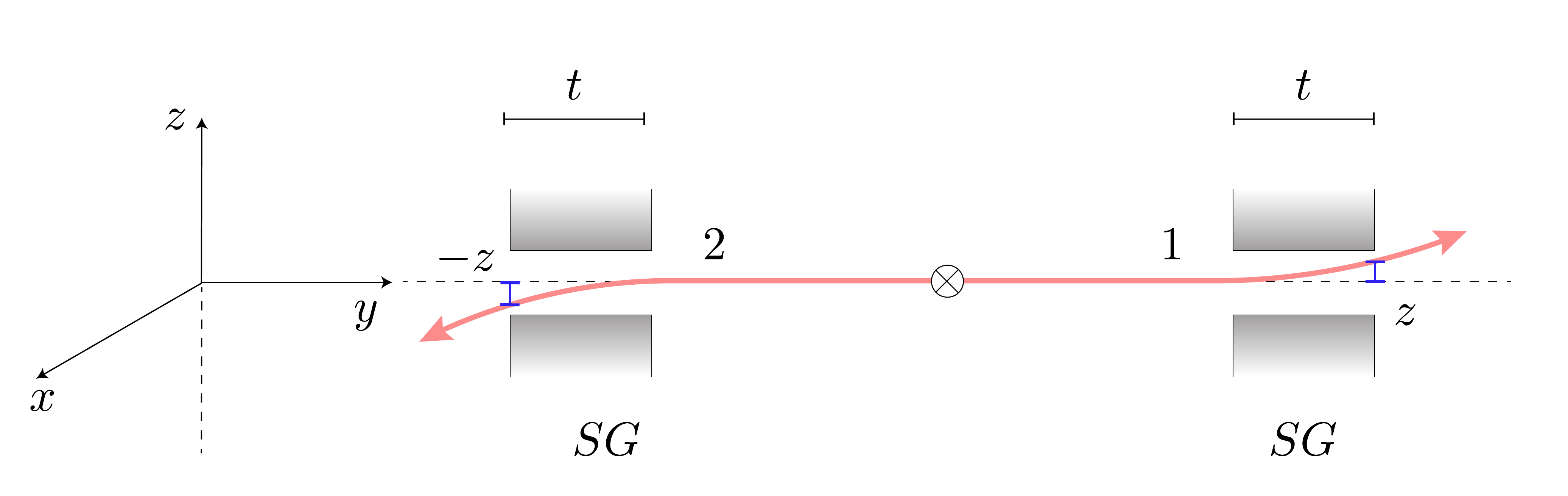}
	\caption{Stern-Gerlach apparata }
	\label{fig:fig-1}
\end{figure}

The total Hilbert space of the system is ${\cal H}=\mathbb{C}^2\otimes\mathbb{C}^2\cong\mathbb{C}^4=\{(z^1,z^2,z^3,z^4), \ z^k\in\mathbb{C}, \  k=1,2,3,4\}$ and clearly $|\psi(t=0)>\in S^7\subset {\cal H}$. After time $t$ the state vector becomes [3] 
\begin{equation}
\begin{split}
|\psi(t,z)>&={{1}\over{\sqrt{2}}}(e^{-i\kappa tz_1}|\uparrow_1>\otimes e^{i\kappa tz_2}|\downarrow_2>-e^{i\kappa tz_1}|\downarrow_1>\otimes e^{-i\kappa tz_2}|\uparrow_2>)\\&\cong{{1}\over{\sqrt{2}}}(e^{-2i\kappa tz}\left(^{1}_{0}\right)_1\otimes\left(^{0}_{1}\right)_2-e^{2i\kappa tz}\left(^{0}_{1}\right)_1\otimes\left(^{1}_{0}\right)_2)\\&\cong
\begin{pmatrix}
0\\
\dfrac{e^{-2i\kappa tz}}{\sqrt{2}}\\
\dfrac{-e^{2i\kappa tz}}{\sqrt{2}}\\
0
\end{pmatrix}
\in S^7,
\end{split}
\end{equation} 
with $\kappa={{\mu\alpha}\over{\hbar}}$ and $-2z=z_2-z_1$ since $z_2=-z_1=-z$ where $z_1$ and $z_2$ are the vertical deviation of the particles in the measurement process. When $t$ varies, so do $z_1$ and $z_2$, and therefore $z$. These changes reflect themselves in $(0,z^2,z^3,0)={{1}\over{\sqrt{2}}}(0,e^{-2i\kappa tz},-e^{2i\kappa tz},0)={{1}\over{\sqrt{2}}}(0,e^{-2i\kappa tz_1},-e^{-2i\kappa tz_2},0)$ which moves along a great circle $S^1$ in $S^7$. It can be shown that if $<v>$ is the average vertical ($z$ direction) velocity of the decaying particles, the period for running through the circle is $\Delta t=\sqrt{{{\pi}\over{\kappa <v>}}}$. In fact, $e^{i2\kappa(t+\Delta t)(z+\Delta z)}=e^{i2\kappa tz}$, with $\Delta z=<v>\Delta t$ implies $\Delta t(<v>t+z+<v>\Delta t)={{\pi}\over{\kappa}}$ which leads to the above $\Delta t$ using the initial condition $(t,z)=(0,0)$. So $S^1\subset S^7$ \textit{plays the role of a bridge} or ``wormhole" between the two non interacting (and therefore not violating causality [8]) correlated (entangled) particles 1 and 2. 

Since $|\psi(t,z)>$ is equivalent to $|\psi(t,z)>^\prime=e^{i\varphi}|\psi(t,z)>\equiv |\psi(t,z;\varphi)>$ with $\varphi\in[0,2\pi)$, the ``real" quantum state of the particles is represented by the great circle in $S^7$
\begin{equation}
\{|\psi(t,z;\varphi)>\}_{\varphi\in[0,2\pi)}\cong
\begin{Bmatrix}
\begin{pmatrix}
0\\
\dfrac{e^{i(\varphi-2\kappa tz)}}{\sqrt{2}}\\
-\dfrac{e^{i(\varphi+2\kappa tz)}}{\sqrt{2}}\\
0
\end{pmatrix}
\end{Bmatrix}
_{\varphi\in[0,2\pi)}\equiv\Psi(t,z),
\end{equation} 
which is nothing but a \textit{point} in $\mathbb{C}P^3$, base space of the $U(1)$-bundle
\begin{equation}
U(1)\to S^7\to \mathbb{C}P^3.
\end{equation}

$\mathbb{C}P^3$ is the space of complex lines through the origin in $\mathbb{C}^4$; each complex line intersects $S^7$ in a unit great circle (3) which is a geodesic [6] of the Fubini-Study metric [7] of $\mathbb{C}P^3$ which, in terms of the affine coordinates $Z^i_{\{n\}}$, $i=1,2,3$, $n=1,2,3,4$ in $\mathbb{C}P^3$ is given by 
\begin{equation}
\begin{split}
{g_{\mathbb{C}P^3}}_{\{n\}}&={g_{\mathbb{C}P^3}}_{\{n\}}^\dagger=\overline{{g_{\mathbb{C}P^3}}_{\{n\}}^T}\equiv({g_{ij}}_{\{n\}}),\\
{g_{ii}}_{\{n\}}&={{1}\over{(A_{\{n\}}})^2}(A_{\{n\}}+|Z^i_{\{n\}}|^2),\\
{g_{ij}}_{\{n\}}&={{1}\over{(A_{\{n\}}})^2}\bar{Z}^i_{\{n\}}Z^j_{\{n\}}, \ i\neq j,
\end{split}
\end{equation}  
where $A_{\{n\}}=1+\Sigma^3_{k=1}|Z^k_{\{n\}}|^2$. $n$ labels the charts on $\mathbb{C}P^3$ ($z^n\neq 0$ in chart $n$) with 
\begin{equation}
\begin{split}
\left[ Z^1_{\{1\}},Z^2_{\{1\}},Z^3_{\{1\}}\right] &=\left[ {{z^2}\over{z^1}},{{z^3}\over{z^1}},{{z^4}\over{z^1}}\right] ,\\
\left[ Z^1_{\{2\}},Z^2_{\{2\}},Z^3_{\{2\}}\right] &=\left[ {{z^1}\over{z^2}},{{z^3}\over{z^2}},{{z^4}\over{z^2}}\right] ,\\
\left[ Z^1_{\{3\}},Z^2_{\{3\}},Z^3_{\{3\}}\right] &=\left[ {{z^1}\over{z^3}},{{z^2}\over{z^3}},{{z^4}\over{z^3}}\right] ,\\
\left[ Z^1_{\{4\}},Z^2_{\{4\}},Z^3_{\{4\}}\right] &=\left[ {{z^1}\over{z^4}},{{z^2}\over{z^4}},{{z^3}\over{z^4}}\right] .
\end{split}
\end{equation}

It is easy to verify that any of the states in $\Psi(t,z)$ is entangled i.e. it has associated with it a non vanishing entanglement entropy (E.E.). In fact, the density operator of the pure state $|\psi(t,z;\varphi)>$ is
\begin{equation}
\begin{split}
\rho&=|\psi(t,z;\varphi)>\otimes<\psi(t,z;\varphi)|\\
&={{1}\over{2}}(e^{i(\varphi-2\kappa tz)}|\uparrow_1>\otimes|\downarrow_2>-e^{i(\varphi+2\kappa tz)}|\downarrow_1>\otimes|\uparrow_2>)\\
&\otimes(e^{-i(\varphi-2\kappa tz)}<\uparrow_1|\otimes<\downarrow_2|-e^{-i(\varphi+2\kappa tz)}<\downarrow_1|\otimes<\uparrow_2|)
\end{split}
\end{equation} 
with reduced density operators 
\begin{equation}
\begin{split}
\rho^{red.}_1&=tr_2\rho=<\uparrow_2|\rho|\uparrow_2>+<\downarrow_2|\rho|\downarrow_2>,\\
\rho^{red.}_2&=tr_1\rho=<\uparrow_1|\rho|\uparrow_1>+<\downarrow_1|\rho|\downarrow_1>,
\end{split}
\end{equation}
which in matrix form are
\begin{equation}
(\rho^{red.}_1)_{ij}=(\rho^{red.}_2)_{ij}=\dfrac{1}{2}
\begin{pmatrix}
1 & 0 \\
0 & 1 \\
\end{pmatrix} .
\end{equation} 
Therefore, the entanglement entropy of $\Psi(t,z)$ is 
\begin{equation}
E.E.(\Psi(t,z))=S(\rho^{red.}_1)=S(\rho^{red.}_2)=2\times(-{{1}\over{2}}ln{{1}\over{2}})=ln2>0.
\end{equation}
\section{$\gamma\gamma$ decays}
The previous analysis repeats almost unmodified for two photon decays in the cases:

i) negative parity, e.g. $e^+e^-\to \gamma\gamma$, with 
\begin{equation}
\psi_-(t=0)={{1}\over{\sqrt{2}}}(\hat{x}_1\otimes\hat{y}_2-\hat{y}_1\otimes\hat{x}_2)\cong
\begin{pmatrix}
0\\
\dfrac{1}{\sqrt{2}}\\
-\dfrac{1}{\sqrt{2}}\\
0
\end{pmatrix}
\in\mathbb{C}^4,
\end{equation}
and 

ii) positive parity, e.g. $J=0\to J=1\to J=0$, 
\begin{equation}
\psi+(t=0)={{1}\over{\sqrt{2}}}(\hat{x}_1\otimes\hat{x}_2+\hat{y}_1\otimes\hat{y}_2)\cong
\begin{pmatrix}
\dfrac{1}{\sqrt{2}}\\
0\\
0\\
\dfrac{1}{\sqrt{2}}
\end{pmatrix}
\in\mathbb{C}^4.
\end{equation}
In both cases,
\begin{equation}
E.E.(\psi_-)=E.E.(\psi_+)=ln2.
\end{equation} 
Again, the bridge between the escaping photons is a great circle in $S^7$ with $U(1)\to S^7\to\mathbb{C}P^3$ the relevant $U(1)$-bundle.

\section{Conclusions}
We show that the unique candidate for playing the role of an Einstein-Rosen bridge in Einstein-Podolsky-Rosen-Bohm two particle decays, is a great unit circle $S^1$ in the 7-sphere. Though this $S^1$ does not live in ordinary spacetime but lies in the complex Hilbert space of the system ($S^1\subset S^7\subset \mathbb{C}^2\times\mathbb{C}^2\cong\mathbb{C}^4\equiv{\cal H}$), it has an explicite relation with the spacetime variables $\{t,z\}$ of the measuring S.G. apparata. The period for running around the circle is $\Delta t=\sqrt{{{\pi}\over{\kappa <v>}}}\simeq 5\times 10^{-4}s$, for $\alpha\simeq 1{{Gauss}\over{m}}$, $<v>\simeq 1{{m}\over{s}}$, and $\mu\simeq 10^{-20}{{erg}\over{Gauss}}$; in this sense, if $\Delta t>t$, $S^1$ is a traversable bridge, in contradistinction with the usual wormholes in the spacetime of black holes in general relativity [9].

\section*{Acknowledgement}
We thank Prof. Ildefonso Castro for clarifying us about geodesics in complex projective 3-space.

\section*{References}

1. J. Maldacena and L. Susskind, \textit{Cool horizons for entangled black holes}, Fortschr. Phys. \textbf{61}, 1-31 (2013).
\\
2. A. Einstein, B. Podolsky, and N. Rosen, \textit{Can Quantum-Mechanical Description of Physical Reality Be Considered Complete?}, Phys. Rev. \textbf{47}, 777-780 (1935).
\\
3. D. Bohm, \textit{Quantum Theory}, Prentice-Hall, N.Y. (1951).
\\
4. A. Peres, \textit{Quantum Theory, Concepts and Methods}, Kluwer Academic Publishers, The Netherlands (1993).
\\
5. A. Aspect, P. Grangier, and G. Roger, \textit{Experimental Tests of Realistic Local Theories via Bell's Theorem}, Phys. Rev. Lett. \textbf{47}, 460-463 (1981).
\\
6. I. Castro and L. Vrancken, \textit{Minimal Lagrangian submanifolds in} $\mathbb{C}P^3$ \textit{and the sinh-Gordon equation}, Result. Math. \textbf{40}, 130-143 (2001).
\\
7. T. Eguchi, P.B. Gilkey, and A.J. Hanson, \textit{Gravitation, Gauge Theories and Differential Geometry}, Phys. Rep. \textbf{66}, No. 6, 213-393 (1980).
\\
8. M. Socolovsky, \textit{EPR, Bell, GHZ, and Hardy theorems, and quantum mechanics}, Rev. Cub. F\'is.\textbf{ 22}, No. 2, 104-118 (2005). 
\\
9. E. Witten, \textit{Light Rays, Singularities, and All That}, arXiv: hep-th/1901.03928.
\\
\end{document}